\documentclass[pre,twocolumn,showpacs,floatfix]{revtex4}
\usepackage{graphicx}
\usepackage{tabularx}
\usepackage{amsmath}
\usepackage{amsfonts}
\usepackage{amssymb}
\begin{document}
\title{Polarization state of the optical near--field}
\author{Ga\"etan L\'ev\^eque, G\'erard Colas des Francs, Christian Girard}
\affiliation{Centre d'Elaboration des Mat\'eriaux et d'Etudes
Structurales (CNRS), 29 rue J. Marvig, F-31055 Toulouse, France}
\author{Jean Claude Weeber}
\affiliation{Laboratoire de Physique de l'Universit\'e de
Bourgogne, Optique Submicronique, 9 avenue A. Savary, F-21078
Dijon, France}
\author{Christoph Meier, C\'ecile Robilliard, Renaud Mathevet, and John Weiner}
\affiliation{LCAR (CNRS), Universit\'e Paul SABATIER - 118 Route
de Narbonne - B\^atiment 3R1-B4 - 31062 Toulouse Cedex 04}
\date{\today}
\begin{abstract}
The polarization state of the optical electromagnetic field lying
several nanometers above complex dielectric--air interfaces
reveals the intricate light--matter interaction that occurs in the
near--field zone.From the experimental point of view, access to
this information is not direct and can only be extracted from an
analysis of the polarization state of the detected light. These
polarization states can be calculated by different numerical
methods well--suited to {\it near--field optics}. In this paper,
we apply two different techniques (Localized Green Function Method
and Differential Theory of Gratings) to separate each polarisation
component associated with both electric and magnetic optical
near--fields produced by nanometer sized objects. A simple dipolar
model is used to achieve insight into the physical origin of the
near--field polarization state. In a second stage, accurate
numerical simulations of field maps complete data produced by
analytical models. We conclude this study by demonstrating the
role played by the near--field polarization in the formation of
the local density of states.

\end{abstract} \maketitle
\section{Introduction}
Light interactions with dielectric or metallic surfaces displaying
well--defined subwavelength--sized structures (natural or
lithographically designed) give rise to unusual optical
effects\cite{Pohl-Courjon:1993,Reddick-Warmack-Ferrell:1989,Fischer-Pohl:1989,Dawson-deFornel-Goudonnet:1994,Goudonnet-Bourillot-Adam-DeFornel-Salomon-Vincent:1995,Krenn:1995,Krenn-Weeber-Dereux-Bourillot-Goudonnet-Schider-Leitner-Aussenegg-Girard:1999,Weeber-Bourillot-Dereux-Chen-Goudonnet-Girard:1996,Weeber-Girard-Dereux-Krenn-Goudonnet:1999,Weeber-Dereux-Girard-Krenn-Goudonnet:1999,Weeber-Dereux-Girard-Colas-Krenn-Goudonnet:2000,Girard-Joachim-Gauthier:2000}.
The recently observed ``light confinement state'' in which the
light field is trapped by individual surface defects, belongs to
this class of
phenomena\cite{Weeber-Bourillot-Dereux-Chen-Goudonnet-Girard:1996}.
Although, with usual dielectric materials (silica for example),
the local near--field intensity variations observed around the
particles (or structures) remains moderate over the optical
spectrum (between 10 to 40 per cent of the incident light
intensity), these variations can nevertheless be easily mapped
with the tip of a Photon Scanning Tunneling Microscope
(PSTM)\cite{Weeber-Bourillot-Dereux-Chen-Goudonnet-Girard:1996}.
The images recorded with this technique reveal dramatic changes
when passing from the TM (transverse magnetic) to the TE
(transverse electric) --polarized modes. In general, TM--polarized
light tends to display larger contrast than TE--polarized light.
>From the experimental point of view, the definition of the polarization
direction of the incident and detected intensity must be defined
with respect to a unique incident plane. About six years ago Van
Hulst and collaborators proposed a clever probe configuration
 devoted to polarization mapping
\cite{Propstra-VanHulst:1995}. These authors performed these
measurements by using a combined PSTM/AFM microscope in which
detection is implemented by a microfabricated silicon--nitride
probe.
>From this technique, polarization contrast is extracted
by changing the polarization directions of both the incident and
the detected light. The main findings gathered in this
work\cite{Propstra-VanHulst:1995}, concern the relative efficiency
of the four excitation--detection possibilities
(TE/TE,TE/TM,TM/TE,and TM/TM)  to record a highly resolved PSTM
image. In particular, the efficiency of the TM/TM acquisition mode
is well reproduced. Although a
complete interpretation of this work requires a realistic
numerical implementation of the combined AFM/PSTM probe tip, we
can obtain useful information by analyzing the near--field
polarization state versus the polarization state of the
illumination mode.

In addition, in closely related contexts, the control of the
near--field polarization state provides an interesting and
versatile tool for generating powerful applications (tunneling
time measurements \cite{Balcou-Dutriaux:1997}, highly--resolved
microscopy and spectroscopy \cite{Pohl-Courjon:1993}, surface
plasmon resonance spectroscopy of molecular adlayers
\cite{Jung-Campbell-Chinowsky-Mar-Yee:1998}, atom optics
\cite{Landragin-Courtois-Labeyrie-Vansteenkiste-Westbrook-Aspect:1996,Esslinger-Weidemuller-Hemmerich-Hansch:1993}.
More precisely, in the field of atom optics and interferometry, one is interested in building diffraction gratings
that can play the role of beam splitters. Several devices have been
 successfully realised, ranging from mechanical transmission gratings to light
standing waves in free space or evanescent for a prism. For a
general review the reader is referred to \cite{interfero}. To circumvent some
theoritical limitations \cite{CHenkel} it has been recently proposed \cite{Balykin} to use
 micrometer sized metallic stripes to shape the evanescent field.
A full near-field, metallic/dielectric approach open obviously new
perspectives. In particular the spacing period is no longer linked to the atomic optical transition and a reflection
stucture cannot clug. More, higher harmonics in the optical evanescent field can be tuned to produce
for example a blazed atomic grating. Nevertheless, the optical potential is
strongly related to the light field polarisation, which is a farther motivation
for the present study.
We will begin our theoretical analysis with a simple dipolar
scheme in which the main experimental parameters (incident angle,
optical index, polarization of the incident light, ...) appear
explicitly\cite{Girard-Courjon:1990,Keller-Xiao-Bozhevolnyi:1992,Keller:1996}.
In a second stage, these results will be completed with an {\it
ab--initio} approach allowing objects of arbitrary shape to be
treated exactly.
\section{Polarization of the light above a single dielectric
particle}
\label{DIPOL}
To illustrate the coupling between a polarized incident wave
and a small spherical object lying on the sample,
we consider the simple dipolar model depicted in Fig.~1.
\begin{figure}[h!]\centering
\includegraphics[width=2.00in,angle=-90]{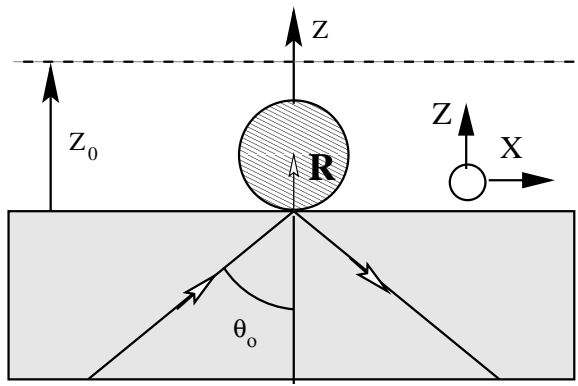}
\caption{Schematic drawing of the model system used in the present
section. A transparent substrate, of optical index $n$ =
$\epsilon^{\frac{1}{2}}$ supports a small dielectric sphere of
radius $R$. The system is illuminated in total internal reflection
with an incident angle $\theta_{0}$ and ${\bf R}$ = $(0,0,R)$. xOz
is the incident plane.} \label{FIG_GEO}
\end{figure}
The substrate modifies the polarizability
$\alpha_{0}(\omega)$ of the particle. We have then
\begin{equation}
\label{eq:alphaef}
 \alpha^{eff}( \mbox{\bf R} ,\omega )=
 \alpha_0(\omega )\cdot \mbox{\bf M}( \mbox{\bf R} ,\omega )
\end{equation}
with
\begin{equation}
\label{eq:matrix}
 \mbox{\bf M}( \mbox{\bf R},\omega )= [ \mbox{{\bf 1}}- \mbox{\bf S}_s(
\mbox{\bf R}, \mbox{\bf R},\omega )\cdot \alpha_0(\omega
)]^{-1}
\end{equation}
where $\mbox{\bf S}_s(\mbox{\bf R} , \mbox{\bf R},\omega )$ is the
nonretarded propagator associated with the bare surface, and
${\mbox{\bf R}}=(0,0,R)$ labels the particle location. Within this
description, the optical properties of the spherical
particle--surface supersystem is described in terms of a
``dressed''
polarizability\cite{Girard-Dereux-Weeber:1998,Keller:1996}. The
analytical form of $\alpha^{eff}$ can be derived from Eq. (10) of
reference 5. This dyadic tensor remains diagonal with two
independent components $\alpha^{eff}_{\perp}$ (perpendicular to the interface) and
$\alpha^{eff}_{\parallel}$ (parallel to the interface):
\begin{equation}
\label{eq:aper}
 \alpha^{eff}_{\parallel} ( \mbox{\bf R} ,\omega )=
\frac{8(n^{2}+1)\alpha_0(\omega)R^{3}}
{8(n^{2}+1)R^{3}-\alpha_0(\omega)(n^{2}-1)}
\end{equation}
and
\begin{equation}
\label{eq:apar}
 \alpha^{eff}_{\perp} ( \mbox{\bf R} ,\omega )=
\frac{4(n^{2}+1)\alpha_0(\omega)R^{3}}
{4(n^{2}+1)R^{3}-\alpha_0(\omega)(n^{2}-1)}
\end{equation}
where $n$ is the optical index of refraction of the substrate.
\subsection{New field components in the near--field domain}
At an observation point ${\bf r}$ located above the sample ($z$
$>$ 0) and in the immediate proximity of the particle, the
incident light field is locally distorted. As illustrated in Ref.
\cite{Martin-Girard-Dereux:1995:2}, these distortions generate not
only a profound modification of the intensity level (both electric
and magnetic), but also a complete change of the polarization
state. At subwavelength distances from the scatterers, we expect
therefore to observe the occurrence of new components that were
absent in the incident field $\{{\bf E}_{0}({\bf r},t); {\bf
B}_{0}({\bf r},t)\}$. The physical origin of this polarization
transfer can be easily understood if we introduce the two relevant
field propagators ${\bf S}_{0}$ and ${\bf Q}_{0}$ that establish
the physical link between the oscillating dipole $\mu(t)$ =
$\alpha^{ef}({\bf R},\omega_0)\cdot{\bf E}_{0}({\bf R},t)$ and the
new near--field state $\{{\bf E}({\bf r},t);{\bf B}({\bf r},t)\}$
generated above the particle
\begin{equation}
\label{eq:Elec}
{\bf E}({\bf r},t)=
{\bf E}_{0}({\bf r},t)+{\bf S}_{0}({\bf r},{\bf R})
\cdot\alpha^{eff}({\bf R},\omega_0)\cdot{\bf E}_{0}({\bf R},t)
\end{equation}
and
\begin{equation}
\label{eq:Mag}
{\bf B}({\bf r},t)=
{\bf B}_{0}({\bf r},t)+{\bf Q}_{0}({\bf r},{\bf R},\omega_0)
\cdot\alpha^{eff}({\bf R},\omega_0)\cdot{\bf E}_{0}({\bf R},t)
\end{equation}
where
in the near--field zone, i.e. when $|{\bf r}-{\bf R}|$ $<$
$\lambda_{0}=2\pi c/\omega_{0}$,
\begin{equation}
\label{eq:Lan1}
{\bf S}_{0}({\bf r},{\bf R})=
\frac{3({\bf r}-{\bf R})({\bf r}-{\bf R})-|{\bf r}-{\bf R}|^{2}{\bf 1}}
{|{\bf r}-{\bf R}|^{5}}
\end{equation}
and
\begin{eqnarray}
\label{eq:Lan2}
\mbox{{\bf Q}}_{0}(\mbox{{\bf r}},\mbox{{\bf R}},\omega_0)
 =  \frac{i\omega_0}{c|\mbox{{\bf r}}-\mbox{{\bf R}}|^3}\left(
\begin{array}{ccc}
0   & -(z-R) & y  \\
z-R & 0      & -x \\
-y  & x      & 0
\end{array}
\right)
\end{eqnarray}
The discussion of the two equations (5) and (6) can be made easier
when considering specific examples.
\subsection{Illumination with a TE--polarized surface wave}
\label{SMODE}
In this incident polarization mode, the electric field
is directed along the (OY) axis. We have then
${\bf E}_{0}({\bf R},t)$ = $(0,E_{0}(t),0)$.
Let us see what happens
when the observation point moves along the diagonal straigth line (A--B)
schematized in figure 2.
\begin{figure}[h!]\centering
\includegraphics[width=3.00in,angle=-90]{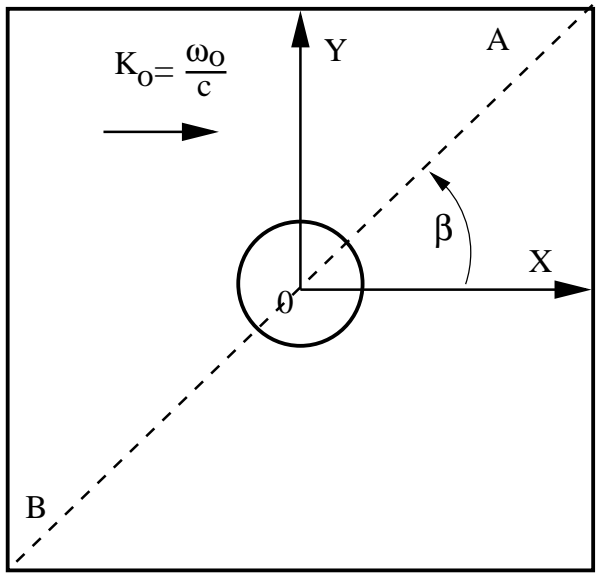} \caption{Top view of the single spherical system depicted
in figure \ref{FIG_GEO}. The straight line (A--B) that passes over
the sphere center at a constant height $z$ is used to evaluate the
polarization change produced by the particle.} \label{TOP_W}
\end{figure}
For a given $\beta$ angle, the introduction of the position vector ${\bf
r}$ = $(r\cos(\beta), r\sin(\beta),z)$ (where $r$ varies between
$-\infty$ and $\infty$) along the line (A--B), allows the electric
field polarization evolution to be observed when passing over the
particle. Each cartesian component can be simply extracted from
Eq. (5). This leads to three analytical relations:
\begin{equation}
E_{x}(t)=E_{0}(t)\frac{\alpha^{eff}_{\parallel}(
\mbox{\bf R},\omega )3r^{2}\cos(\beta)\sin(\beta)}
{|r^{2}+(z-R)^{2}|^{5/2}}
\end{equation}
\begin{equation}
E_{y}(t)=E_{0}(t)\{1+\alpha^{eff}_{\parallel}(\mbox{\bf R},\omega_{0})
{\cal T}(\beta)\},
\end{equation}
with
\begin{equation}
{\cal T}(\beta)=\frac{2r^{2}\sin^{2}(\beta)-r^{2}\cos^{2}(\beta)-
(z-R)^{2}}{|r^{2}+(z-R)^{2}|^{5/2}},
\end{equation}
and
\begin{equation}
E_{z}(t)=E_{0}(t)\frac{\alpha^{eff}_{\parallel}(
\mbox{\bf R},\omega_{0})3r\sin(\beta)(z-R)}
{|r^{2}+(z-R)^{2}|^{5/2}}.
\end{equation}
Some interesting features can be deduced from these equations. (i)
First, we observe that the polarisation is not modified when the
observation point is perpendicular the particle center (i.e. when
r = 0). The set of equations reduces then to
\begin{equation}
E_{x}(t)=0,
\end{equation}
\begin{equation}
E_{y}(t)=E_{0}(t)
\{1-\frac{\alpha^{eff}_{\parallel}(0,0,z,\omega_0)}{|z-R|^{3}}\},
\end{equation}
and
\begin{equation}
E_{z}(t)=0.
\end{equation}
Clearly the effective polarizability reduces the electric field
magnitude compared to its initial value. This fact explains the
observation of contrasted dark zones above small particles lighted
with the TE polarization (c.f. Fig. (4)). (ii) Around the particle (when R/2 $<$ r
$<$ 2R) two new components, namely $E_{x}(t)$ and $E_{z}(t)$,
define a new local polarization state. These components vanish
again when the observation point moves away from the particle.
Similar relations can be derived for the TM--polarized mode from
Eq. (5).

To conclude this section, let us examine what happens with the
magnetic field part (cf. Eq. 6). Since in the reference system of
Fig. (1), the incident magnetic field displays two components
different from zero, ${\bf B}_{0}({\bf r},t)$ =
$(B_{0x}(t),0,B_{0z}(t))$, we can write
\begin{equation}
B_{x}(t)=B_{0x}(t)-\frac{i\omega_{0}(z-R)}{c[r^{2}+(z-R)^{2}]^{3/2}}
\alpha_{\parallel}^{eff}E_{0}(t),
\end{equation}
\begin{equation}
B_{y}(t)=0,
\end{equation}
and
\begin{equation}
B_{z}(t)=B_{0z}(t)+\frac{i\omega_{0}r\cos(\beta)}{c[r^{2}+(z-R)^{2}]^{3/2}}
\alpha_{\parallel}^{eff}E_{0}(t).
\end{equation}
Unlike what happens with the electric field, the particle does not
produce new magnetic field components in the near-field. In this
case, the polarization change corresponds to a different balance
in the initial components. It is important to recall that we have
used the dipole approximation to describe the particle-field
interaction, i.e. the size of the particle is assumed to be small
compared to the wavelength of light. In a more realistic
calculation, with nanostructures of characteristic dimension
$\approx$ 100 nanometers this result is not rigorously exact.
However, we still expect, in TE polarization mode, a negligible
particle contribution to $B_y$ compared to the total magnetic
field intensity.
\section{{\it AB INITIO} study of the near--field
polarization state}
Analytical results presented in the previous section supply
qualitative information about the spatial polarization state
distribution. In a recent paper, analysis of polarization effects
was proposed in the context of near--field optics in which a
limited number of single particles were
investigated\cite{Richard:2001}. Since in many practical
situations experimentalists are interested in lithographically
designed structures, these preliminary analyses must be completed
by {\it ab initio} procedures for solving Maxwell's equations.
\subsection{Localized objects}
\label{LOCALIZED} Recently, theoretical modelling in the vicinity
of localized objects was performed in the framework of the field
susceptibility
method\cite{Girard-Bouju-Dereux:1993,Greffet-Carminati:1997}.
Today, this method is one of the most versatile and reliable
numerical techniques to solve the full set of Maxwell equations
for the typical parameters of near--field optics. It works well
even for metallic nanostructures (see for example
references\cite{Krenn-Weeber-Dereux-Bourillot-Goudonnet-Schider-Leitner-Aussenegg-Girard:1999,Weeber-Girard-Dereux-Krenn-Goudonnet:1999,Weeber-Dereux-Girard-Krenn-Goudonnet:1999}).
This approach (called the Direct Space Integral Equation Method
(DSIEM)) is based on the knowledge of the  retarded dyadic tensor
${\bf S}({\bf r},{\bf r'},\omega)$ associated with a reference
system which, in our problem, is a flat silica
surface\cite{Martin-Piller:1998,Martin-Girard-Smith-Schultz:1999}.
The numerical procedure considers any object deposited on the
surface as a localized perturbation which is discretized in direct
space over a predefined volume mesh of N points $\{{\bf R_{i}}\}$.
In a first step, the electric field distribution ${\bf E}({\bf
R_{i}},\omega)$ is determined self-consistently inside the
perturbations (i.e., the source field). At this stage, a
renormalization procedure associated to the depolarization effect
is applied to take care of the self-interaction of each
discretization cell. The final step relies on the Huygens--Fresnel
principle to compute the electromagnetic field ${\bf E}({\bf
r},\omega)$ on the basis of the knowledge of the field inside the
localized perturbations ${\bf E}({\bf R_{i}},\omega)$. The two
main computational steps can be summarized as follows:
\par
\bigskip
\noindent
{\it (i) Local field computation inside the source field}
\par
\bigskip
\noindent
\begin{equation}
{\bf E}({\bf R}_{i},\omega)=
\sum_{j}{\cal K}({\bf R}_{i},{\bf R}_{j},\omega)\cdot{\bf E}_{0}
({\bf R}_{j},\omega),
\end{equation}
where ${\cal K}$ labels the generalized field propagator of the entire
system (localized object plus bare silica surface).
In the $\{{\bf R}_{i};{\bf R}_{j}\}$ representation
it is given by
\begin{equation}
{\cal K}({\bf R}_{i},{\bf R}_{j},\omega)
=\delta_{i,j}+v_{j}{\cal S}({\bf R}_{i},{\bf R}_{j},\omega)
\cdot \chi({\bf R}_{j},\omega),
\end{equation}
where $\chi$ represents the electric susceptibility
of the localized object, $v_j$ is the volume of the elementary discretization cell, and ${\cal S}$ is the field--susceptibility of the entire system.
This last quantity is usually computed by solving Dyson's equation:
\begin{eqnarray}
{\cal S}({\bf R}_{i},{\bf R}_{j},\omega)=
{\bf S}({\bf R}_{i},{\bf R}_{j},\omega)+
\\
\nonumber
\sum_{k}v_{k}{\bf S}({\bf R}_{i},{\bf R}_{k},\omega)\cdot
\chi({\bf R}_{k},\omega)\cdot{\cal S}({\bf R}_{k},{\bf R}_{j},\omega),
\end{eqnarray}
{\it (ii) Electric and magnetic near--field mapping computation
around the source field region}
\par
\bigskip
\noindent
\begin{eqnarray}
{\bf E}({\bf r},\omega)=
{\bf E}_{0}({\bf r},\omega)+
\\
\nonumber
\sum_{i}v_{i}{\bf S}({\bf r},{\bf R}_{i},\omega)
\cdot\chi({\bf R}_{i},\omega)\cdot{\bf E}({\bf R}_{i},\omega).
\end{eqnarray}
and
\begin{eqnarray}
{\bf B}({\bf r},\omega)=
{\bf B}_{0}({\bf r},\omega)+
\\
\nonumber
\sum_{i}v_{i}{\bf Q}({\bf r},{\bf R}_{i},\omega)
\cdot\chi({\bf R}_{i},\omega)\cdot{\bf E}({\bf R}_{i},\omega).
\end{eqnarray}
In the numerical work to be discussed in this section,
the retarded propagators ${\bf S}$ and ${\bf Q}$
have been chosen in reference\cite{Girard-Weeber-Dereux-Martin-Goudonnet:1997}.

The test--object we consider in this section is
the word {\bf OPTICS} engraved at the surface of a $Ti0_{2}$
layer deposited on a silica surface.
\begin{figure}[h!]\centering
\includegraphics[width=3.25in,angle=-90]{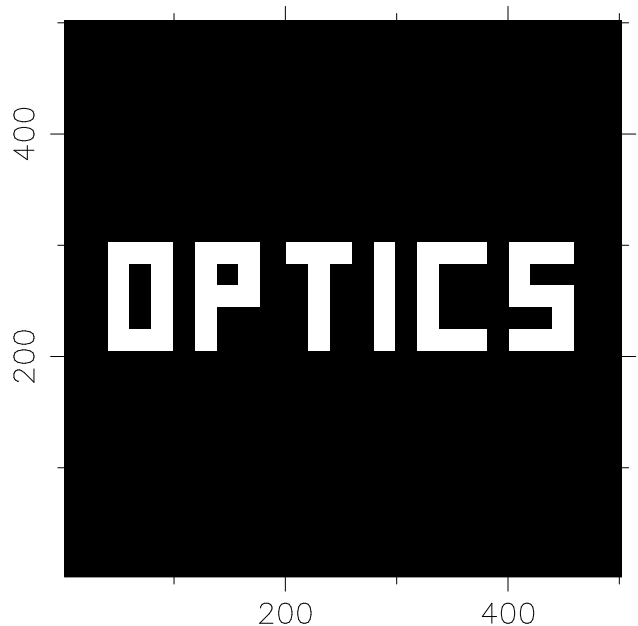} \caption{Top view
of the pattern used in the simulations of the {\it{ab initio}}
studies. The computational window is 500 $\times$ 500 nm$^{2}$. }
\label{OPTICS}
\end{figure}
Intentionally we have chosen a planar structure devoid of any
symmetry. In order to gain more insight in the polarization
changes occurring around such complex lithographically designed
nanostructures, we analyze in Figs. (4)-(7) the electric and
magnetic near--field intensities generated by each cartesian
component ($E_{x}^{2}$, $E_{y}^{2}$, $E_{z}^{2}$) and
($B_{x}^{2}$, $B_{y}^{2}$, $B_{z}^{2}$). For comparison, the
square modulii are also provided.
\subsubsection{Dielectric materials}
The lateral dimensions of the object are given in Fig.
(\ref{OPTICS}). The thickness and the optical index of the $TiO_2$ pattern
is 20 nm and 2.1, respectively. The
wavelength of the incident laser is 633 nm. All fields are
computed 10 nm above the surface of the structure, i.e. 30 nm
above the glass-air interface. \label{DM}
\begin{figure}[h!]\centering
\includegraphics[width=3.25in,angle=-90]{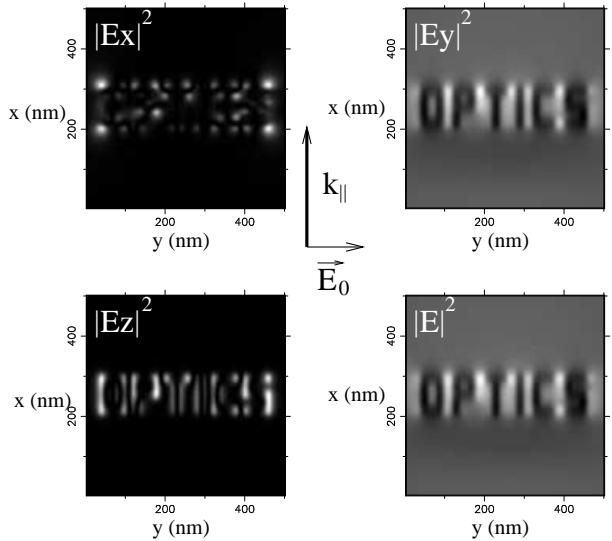}
\caption{Grey scale representation of the electric field
distribution computed above the topographical object depicted in
Fig.3. The calculation is performed in the TE--polarized mode and
the arrow indicates the propagation direction of the surface wave.
Extreme values of the components of the electric field (normalized
by the incident field) are : 0.000 (min) and 0.153E-1 (max) for
$E_x^2$, 0.674 and 1.67 for $E_y^2$, 0.000 and 0.747E-1 for
$E_z^2$, 0.681 and 1.68 for $E^2$. } \label{E_TE}
\end{figure}
\begin{figure}[h!]\centering
\includegraphics[width=3.25in,angle=-90]{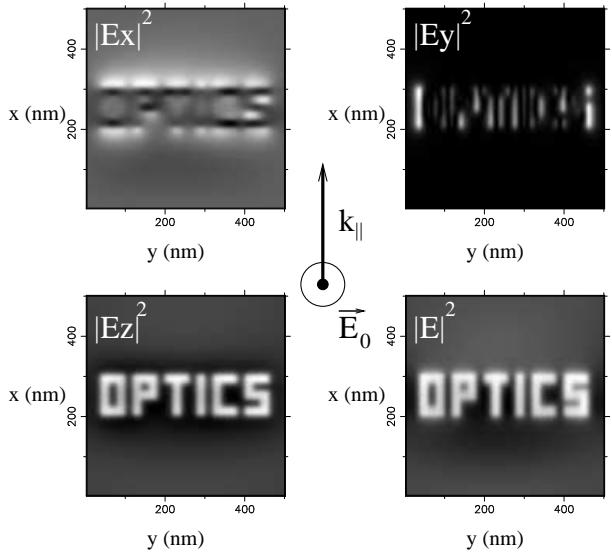}
\caption{Same as figure 4, but in TM--polarized mode. Extreme
values are : 0.206 (min) and 0.436 (max) for $E_x^2$, 0.700E-9 and
0.216E-1 for $E_y^2$, 0.613 and 1.06 for $E_z^2$, 0.908 and 1.31
for $E^2$. } \label{E_TM}
\end{figure}
The incident light is a TM/TE--polarized evanescent surface wave
traveling along the $Ox$ axis. This illumination condition is used
in the Photon Scanning Tunneling Microscope (PSTM). Some general
comments can be made about these results.
First, all components of both the electric and magnetic fields
have been excited in the near-zone. The occurence of these new
components is a pure near--field effect because it is always
localized around the structures. In Figs. (\ref{E_TE}) and
(\ref{E_TM}) we display the electric field part. As predicted in
section 2, we recover the appearance of two additional
components, $E_{x}$ and $E_{z}$, when the object is excited by a
TE--polarized surface wave. In agreement with the PSTM results,
numerous regions appear with a dark contrast
and a moderate intensity level.

As expected, the excellent image--object
relation currently observed in the TM--polarized mode is mainly
provided by the field component $E_{z}$ normal to the object. The
two other contributions tend to slightly degrade the total pattern
$E^{2}$ composed by the superposition of the three maps
$E_{x}^{2}$, $E_{y}^{2}$ and $E_{z}^{2}$.

The magnetic near--field intensity maps (cf. Figs. 6, 7) also show
a significant confinement of the magnetic field over the particle
which reverses the contrast with respect to the electric map.
Similarly to what happens with the electric field, the role played
by the additional components can degrade the topographic
information contained in the complete field maps. Notice in
Fig.\,(6) that, as mentioned in section \ref{DIPOL}, the new
y-component of the magnetic field is very small compared to the
total magnetic field.
\begin{figure}[h!]\centering
\includegraphics[width=3.25in,angle=-90]{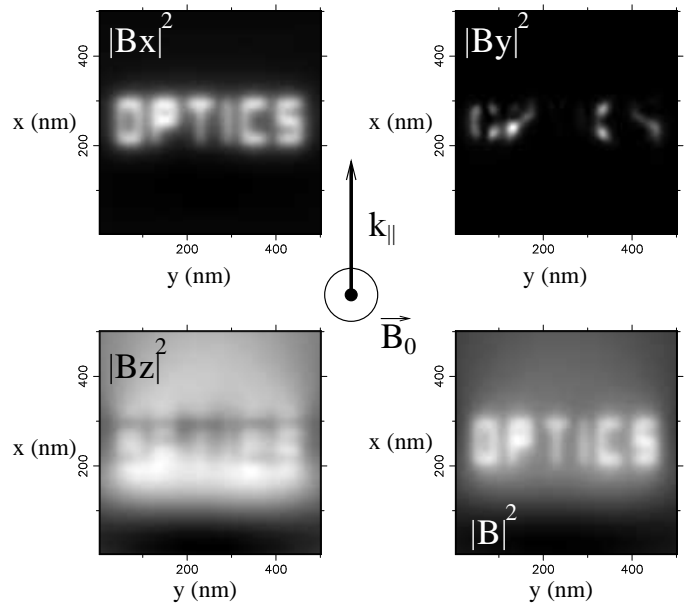}
\caption{ Grey scale representation of the magnetic field
distribution computed above the topographical object depicted in
figure 3. The calculation is performed in the TE--polarized mode.
Each map is normalized with respect to the incident magnetic field
intensity. Extreme values are : 0.287 (min) and 0.396 (max) for
$B_x^2$, 0.696E-13 and 0.490E-4 for $B_y^2$, 0.684 and 0.772 for
$B_z^2$, 0.973 and 1.145 for $B^2$. } \label{B_TE}
\end{figure}
\begin{figure}[h!]\centering
\includegraphics[width=3.25in,angle=-90]{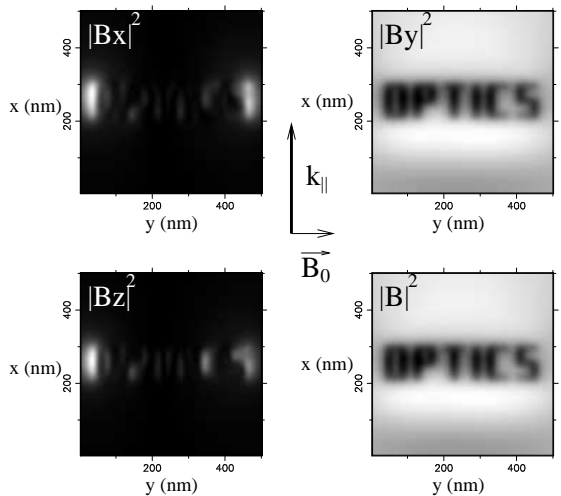}
\caption{ Same as figure 6, but in TM--polarized mode. Extreme
values are : 0.109E-7 (min) and 0.856E-2 (max) for $B_x^2$, 0.710
and 1.11 for $B_y^2$, 0.107E-7 and 0.190E-1 for $B_z^2$, 0.711 and
1.11 for $B^2$. } \label{B_TM}
\end{figure}
\subsubsection{Metallic materials}
\label{MM} In the above formalism, the only parameter distinguishing
metallic from dielectric objects is the linear susceptibility
$\chi(\mbox{\bf{r}}',\omega)$. Alternative procedures can be
adopted to describe the metallic susceptibility. For example, a
direct route would consist in expanding the susceptibility in a
multipolar series around the geometrical center of the metallic
particle. While this scheme allows non--local and quantum size
effects to be included, it is nevertheless restricted to simple
particle shapes (spheres, ell0ipsoids, etc.).  When dealing with
spherical metallic clusters having a typical radius below 15 nm
this description is mandatory and can be easily included in the
DSIEM formalism\cite{Girard:1992}. For the applications discussed
in this paper, involving lithographically designed metallic
structures larger than this critical size, we can adopt the
discretization of $\chi(\mbox{\bf{r}}',\omega)$ over all the
volume occupied by the particle.  In this case, the local
susceptibility is just related to the metal optical index $n$ by
the relation\cite{Weeber-Girard-Dereux-Krenn-Goudonnet:1999}:
\begin{equation}
\chi(\mbox{\bf{r}}',\omega)=
\frac{(n^{2}(\omega)-1)}{4\pi}
\end{equation}
In the visible range, the numerical data for describing both real
and imaginary parts of $n$ have been tabulated by
Palik\cite{Palik:1985} for different metals.
\begin{figure}[h!]\centering
\includegraphics[width=3.25in,angle=-90]{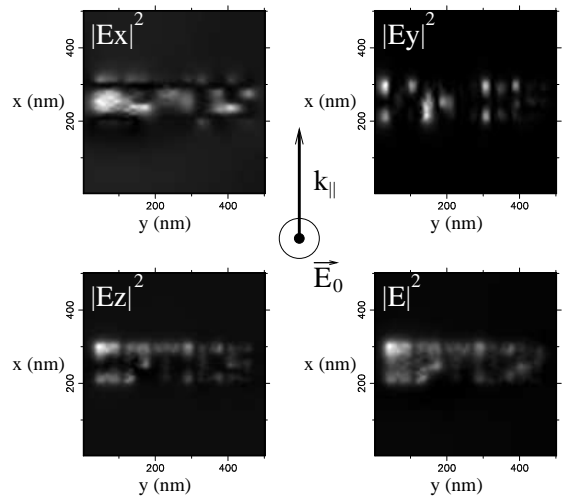}
\caption{Grey scale representation of the electric field
distribution computed above the topographical object depicted in
Fig. 3. In this application the supported structure is metallic
(gold). The incident wavelength is 760 nm. The calculation is
performed in the plane $z_{0}$ = 30 nm in the TM--polarized mode
and the arrow indicates the propagation direction of the surface
wave. Extremal values of the components of the normalized electric
field are : 0.118E-2 (min) and 3.13 (max) for $E_x^2$, 0.146E-5
and 1.72 for $E_y^2$, 0.237 and 10.1 for $E_z^2$, 0.851 and 10.6
for $E^2$. } \label{E_TM_GOLD}
\end{figure}
We present in Fig. (\ref{E_TM_GOLD}) a gray scale representation
of the electric field distribution computed above the
topographical object depicted in Fig. 3. In this case, the high
optical metal index generates complex field patterns without clear
relation to the topography. Furthermore the possible excitation of
localized plasmons reinforces this phenomena and some parts of the
localized metal pattern (e.g. the corners) can even behave as an
efficient light sources.
\subsection{Periodic surface structures}
When working with periodic surface structures, the localized
Green's function method described above is no longer applicable.
But any figure can be decomposed in points (direct space) or Fourier components (
reciproqual space). We thus can use two methods wether the real object contain few
 points or few Fourier components. In the case of periodic gratings, the field distribution
 can be investigated with this second class of methods
\cite{Petit:1980,Maystre-Neviere:1978,Montiel-Neviere:1994}. The
so--called differential theory of gratings (DTG) was originally
developed twenty years ago to predict the efficiencies of one--
and two--dimensional diffraction gratings. Based on a rigorous
treatment of Maxwell's equations, this method can also be used
efficiently to determine the optical near--field scattered by
three dimensional periodic objects. In the following subsection,
in order to avoid a complete presentation of this
well--established technique, we will only summarize the essential
steps of the computational procedure.

As in previous sections, we are interested in the electromagnetic
near-field diffracted above objects engraved on an interface
illuminated by total internal reflection. When using the DTG
method\cite{Petit:1980}, the electromagnetic field above the
grating can be expanded in a Fourier series
\begin{equation}
\label{DIFFRACTED}
\mbox{\bf{A}}({\bf r}) = \sum_{p = -\infty}^{ + \infty}\sum_{q
= -\infty}^{ + \infty}\;
\mbox{\bf{A}}_{p,q}e^{i\gamma_{p,q}z}e^{i\mbox{\bf{k}}_{\parallel p,q}\cdot
\mbox{\bf{l}}} \:,
\end{equation}
where ${\bf r}$ = $({\bf{l}},z)$ = $(x,y,z)$, ${\bf{A}}({\bf r})$
represents either the electric field ${\bf{E}}({\bf r})$ or the
magnetic field ${\bf{B}}({\bf r})$. The 3D--wave vectors
$\mbox{\bf{k}}_{p,q} = \left( \mbox{\bf{k}}_{\parallel
p,q},\gamma_{p,q} \right)$, associated with the harmonic $(p,q)$
obey the well-known dispersion equation
\begin{equation}
\label{KPAR}
{\mbox{\bf{k}}_{\parallel p,q}}^2 + \gamma_{p,q}^2 =  n^{2}{{k_0}}^2,
\end{equation}
The set of wave vector $\mbox{\bf{k}}_{\parallel p,q}$ parallel to
the surface are simply defined for each couple of integer numbers
$(p,q)$ by
\begin{equation}
\label{EYBY}
\mbox{\bf{k}}_{\parallel p,q} = \left( nk_{0x} +
p\frac{2\pi}{d_x}\right) \mbox{\bf{u}}_{x} + \left( nk_{0y} +
q\frac{2\pi}{d_y}\right)\mbox{\bf{u}}_{y} \:,
\end{equation}
where $d_x$ and $d_y$ denote respectively the period of the
grating along the 0x-- and 0y--directions.  From Eq. (\ref{KPAR}),
it may be seen that the coefficient $\gamma_{p,q}$ may be either
real or purely imaginary. Real values of $\gamma_{p,q}$
correspond to radiative harmonics while imaginary values introduce
evanescent components in the expansion (\ref{DIFFRACTED}).

In a general way, the six components of the electromagnetic field
$\mbox{\bf{A}}({\bf r})$ can be deduced from two independent
parameters usually named {\it the principal components}. Let us
choose, for example, the y--components $E_y({\bf r})$ and
$B_y({\bf r})$ as {\it principal components}.  It is a simple
matter to show that the Fourier y--components of the field just
above the surface of objects can be expressed as a linear
combination of the y--components of the incident field:
\begin{equation}
\label{coucou}
\left \lbrace
\begin{array}{l}
E_{y p,q}  =
{\mathcal{T}}^{EE}_{pq}\;E_{0y} + {\mathcal{T}}^{EB}_{pq}\;B_{0y} \:,\\
B_{y p,q}  =
{\mathcal{T}}^{BE}_{pq}\;E_{0y} + {\mathcal{T}}^{BB}_{pq}\;B_{0y} \:.
\end{array}
\right.
\end{equation}
The transmission coefficients ${\mathcal{T}}^{E E}$,
${\mathcal{T}}^{E B}$, ${\mathcal{T}}^{B E}$ and
${\mathcal{T}}^{B B}$ describe the coupling between the electric and
magnetic harmonics composing the scattered and the incident field.
These coefficients depend both on the geometry of the sample and on
the angular conditions of incidence but not on the polarization of the
incident light. The polarization of the incident plane wave is
controlled by the values of $B_{0y}$ and $E_{0y}$. From a numerical point of view, the transmission coefficients are obtained by the inversion of a complex square matrix whose dimension is
$2N_{T}$ $\times$ $2N_{T}$ (where $N_{T}$ is the total number of
harmonics used to describe the scattered field in Eq.~(\ref{DIFFRACTED})).
Columns of this matrix contain the Fourier y--components of the electromagnetic field which would have illuminated the periodical objects in order to obtain a pure harmonic field $(p,q)$ just above the nanostructure. A detailed description of the calculation of the matrix elements can be found in Refs.~\onlinecite{Petit:1980,Montiel-Neviere:1994}. With Eq.~(\ref{coucou}), we can calculate all the Fourier components of electric and magnetic fields just above the objects, which are used as initial conditions to obtain the field anywhere.
\begin{figure} \centering
\includegraphics[width=3.25in,angle=-90]{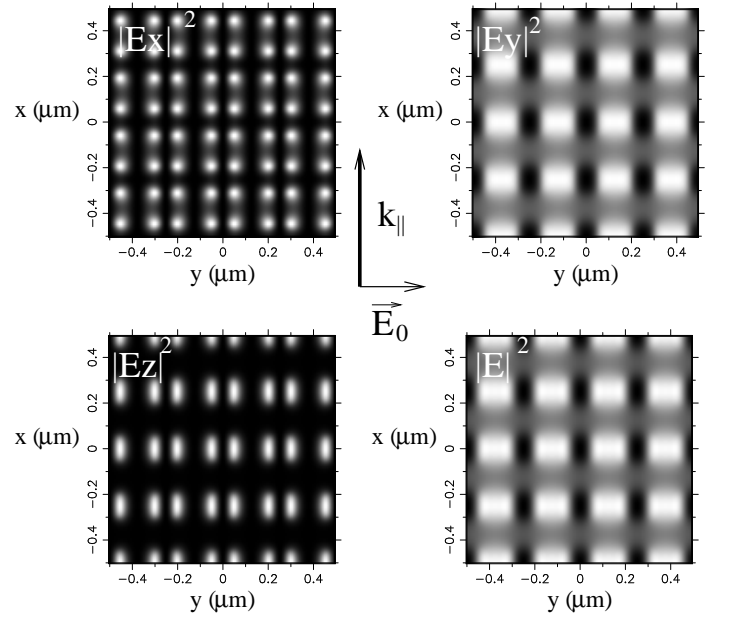}
\caption{Grey scale representation of the electric field
distribution computed above a regular lattice of square shaped
dielectric pads. The calculation is performed in the plane $z_{0}$
= 50 nm in the TE--polarized mode. and the arrow indicates the
propagation direction of the surface wave.  Extremal values of the
components of the normalized electric field are : 0.43E-8 (min)
and 0.875E-2 (max) for $E_x^2$, 0.266 and 0.713 for $E_y^2$,
0.141E-7 and 0.644E-1 for $E_z^2$, 0.275 and 0.741 for $E^2$. }
\label{E_TE_LATTICE}
\end{figure}
\begin{figure}[h!] \centering
\includegraphics[width=3.25in,angle=-90]{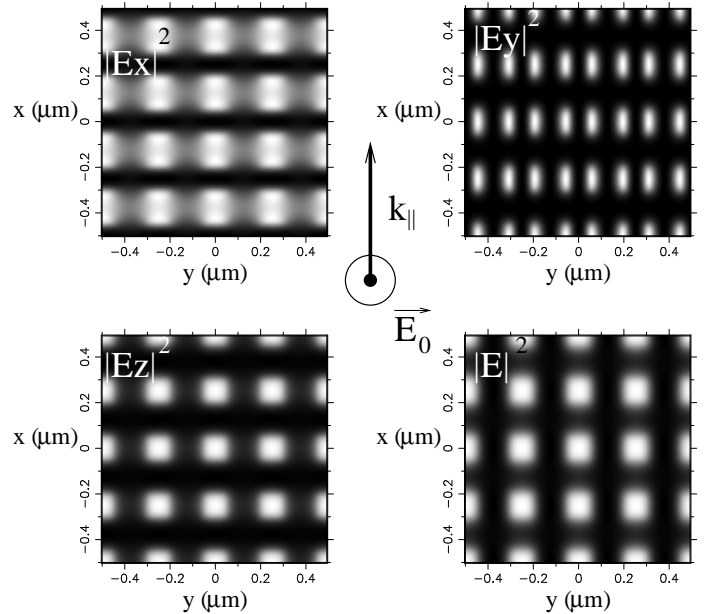}
\caption{ Same as figure (9), but in the TM--polarized mode.
Extremal values of the components of the normalized electric field
are : 0.153 (min) and 0.454 (max) for $E_x^2$, 0.26E-8 and
0.458E-1 for $E_y^2$, 0.513 and 1.49 for $E_z^2$, 0.802 and 1.58
for $E^2$.} \label{E_TM_LATTICE}
\end{figure}
General remarks about contrast, relative intensities and image-object
relation have been made previously (see sections \ref{SMODE}
and \ref{DM}), therefore we only highlight the specific electromagnetic
properties of periodic structures.
\par
We studied a lattice of $100 \times 100 \times 100$ nm$^3$ $TiO_2$ dots,
separated by $150$ nanometers. A strong localization of the electric field
 appears above the pads in TM-polarized mode or between the pads in TE-polarized
mode. Moreover, a careful analysis of the different components
shows that it is possible to create a particular field map such as
field lines oriented along y-axis ($E_{y}$ in TE-mode) or x-axis
($E_{x}$ in TM-mode), or periodic field spots with very different
characteristics (spot size, periodicity and shape) considering the
other components in the two polarization modes. These particular
field components distribution could have a great interest for the
interaction of cold atoms with optical evanescent waves
\cite{Kobayashi-Sangu-Ito-Ohtsu:2000,Birkl-Buchkremer-Dumke-Ertmer:2001}.

\section{Local density of state and polarization effects}
\label{LDOS}
\begin{figure}[h!]\centering
\includegraphics[width=3.00in,angle=0]{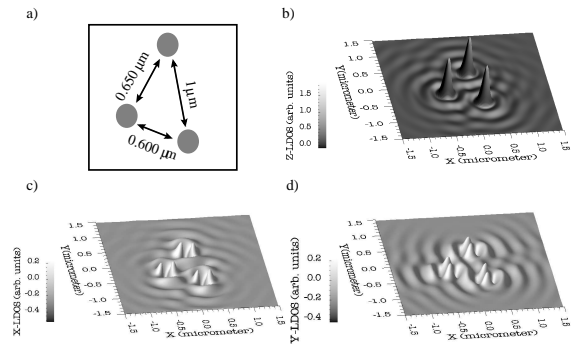}
\caption{a)Top view of the pattern used in the simulations of
section \ref{LDOS} consisting of three cylinders of 100 nm
diameter and 60 nm height. b) z-ldos, c) x-ldos and d) y-ldos 150
nm above the surface at the wavelength $\lambda=2 \pi/ \omega=630$
nm.} \label{ldosfig}
\end{figure}

Unlike what happens with electronic surface states, the Local
Density of photonic States (the so--called photonic LDOS) contains
information related to the polarization of the excitation field.
It is well established that the density of states near surfaces
plays a significant role in near--field physics
\cite{Girard-Joachim-Gauthier:2000}. In particular, the photonic
LDOS is a useful concept for the interpretation of fluorescence
decay rate in the very near--field \cite{Barnes:1998} and could
help in understanding image formation produced by
illuminating--probe SNOM \cite{Dereux-Girard-Weeber:2000}. By
referring to the electric field, we deduce the optical LDOS from
the field--susceptibility of the entire system (plane surface plus
supported nanostructures, see section \ref{LOCALIZED})
\cite{Agarwal:1975c,Dereux-Girard-Weeber:2000}
\begin{equation}
\rho ({\bf r},\omega )=\frac{1}{2\pi ^{2}\omega }
Im\left[
Tr{\cal S}({\bf r},{\bf r},\omega )
\right]\, .
\end{equation}
In this expression, the optical LDOS is related to the square
modulus of the electric field associated with all electromagnetic
eigenmodes of angular frequency $\omega$. Because of the vectorial
character of electromagnetic fields, it is very useful to
introduce three {\it polarized} optical LDOS also called partial
LDOS, so that \cite{Dereux-Girard-Weeber:2000}
\begin{eqnarray}
\rho ({\bf r},\omega) &=& \rho_{x}({\bf r},\omega)+
\rho_{y}({\bf r},\omega)+
\rho_{z}({\bf r},\omega), \\
\rho_{i}({\bf r},\omega) &=& \frac{1}{2\pi ^{2}\omega }
Im {\cal S}_{ii}({\bf r},{\bf r},\omega ),
\qquad i=x,y,z
\end{eqnarray}
The three different {\it polarized} optical LDOS computed over a pattern made
of three dielectric cylinders of optical index 2.1 are represented on
figure \ref{ldosfig}.
At this stage, we can made some
general remark. In the configuration investigated in figure (\ref{ldosfig})
we can verify that both X--LDOS and Y--LDOS are almost
identical with just a rotation of $\pi/2$ between them.
Nevertheless, this properties does not subsist any more
if we reinforce the optical coupling between the dielectric posts.
More precisely, certain arrangements of nanoscale pillars (circular or
ellipsoidal) can force
light waves into states generated by the collective coupling
between the pillars leading to significant difference between
X--LDOS and Y--LDOS maps.

Moreover, although
optical LDOS characterizes the spectroscopic properties of
an electromagnetic system independently of the illumination mode,
it is interesting to note the strong analogy between the dark contrast
over the pads obtained for Y--LDOS and the dark contrast that appears
in the case of s-polarized illumination mode (i-e incident electric field
along the y-axis) observed in section \ref{SMODE}.
In addition, as in the case of optical near--field maps discussed
in the previous section, the {\it polarized} LDOS's
can display significant discrepancies relatively to the shapes of the
underlying objects.

Finally, let us note that partial LDOS's are not only
a powerful mathematical tool but can easily be linked to the physical properties of
electromagnetic systems. The most famous example is the fluorescence
lifetime $\tau$ of a molecule near an interface which critically depends on the
spatial LDOS variation
\cite{Wylie-Sipe:1984,Girard-Martin-Dereux:1995,Metiu:1984}

\section{Conclusion}
On the basis of both  simple analytical model and
sophisticated 3D Maxwell's equations solvers this paper
has focussed on the unusual behaviour of the light polarization
in the near--field.
When  subwavelength patterned objects are
excited by a surface wave of well--defined polarization,
a complex rotation of the light polarization state
can be expected in the near zone. This phenomenon
localized around the scatterer
is a typical near--field effect.
The occurrence of new components is more
pronounced in the electric field than in the magnetic part.
Subwavelength features are present in all components
but with very different energy levels.


\begin{thebibliography}{10}

\bibitem{Pohl-Courjon:1993}
 in {\em Near-field optics}, Vol.~E 242 of {\em NATO ASI}, NATO, edited by D.
  Pohl and D. Courjon (Klu\-wer, Dor\-drecht, 1993).

\bibitem{Reddick-Warmack-Ferrell:1989}
R.~C. Reddick, R.~J. Warmack, and T.~L. Ferrell, Phys. Rev. B {\bf 39},  767
  (1989).

\bibitem{Fischer-Pohl:1989}
U. Fischer and D.W. Pohl, Phys. Rev. Lett. {\bf 62},  458  (1989).

\bibitem{Dawson-deFornel-Goudonnet:1994}
P. Dawson, F. de~Fornel, and J.~P. Goudonnet, Phys. Rev. Lett. {\bf 72},  2927
  (1994).

\bibitem{Goudonnet-Bourillot-Adam-DeFornel-Salomon-Vincent:1995}
J.~P. Goudonnet {\it et~al.}, J. Opt. Am . Soc. {\bf 12},  1749  (1995).

\bibitem{Krenn:1995}
J. Krenn,  in {\em Photons and local probes}, Vol. E 300 of {\em NATO ASI}, edited by
  O. Marti (Klu\-wer, Dor\-drecht, 1995), 181

\bibitem{Krenn-Weeber-Dereux-Bourillot-Goudonnet-Schider-Leitner-Aussenegg-Gir%
ard:1999}
J. Krenn {\it et~al.}, Phys. Rev. B {\bf 60},  5029  (1999).

\bibitem{Weeber-Bourillot-Dereux-Chen-Goudonnet-Girard:1996}
J. Weeber {\it et~al.}, Phys. Rev. Let. {\bf 77},  5332  (1996).

\bibitem{Weeber-Girard-Dereux-Krenn-Goudonnet:1999}
J. Weeber {\it et~al.}, J. Appl. Phys. {\bf 86},  2576  (1999).

\bibitem{Weeber-Dereux-Girard-Krenn-Goudonnet:1999}
J. Weeber {\it et~al.}, Phys. Rev. B {\bf 60},  9061  (1999).

\bibitem{Weeber-Dereux-Girard-Colas-Krenn-Goudonnet:2000}
J. Weeber {\it et~al.}, Phys. Rev. E {\bf 62},  7381  (2000).

\bibitem{Girard-Joachim-Gauthier:2000}
C. Girard, C. Joachim, and S. Gauthier, Rep. Prog. Phys. {\bf 63},  893
  (2000).

\bibitem{Propstra-VanHulst:1995}
K. Propstra and N.~K.~V. Hulst, J. of Microscopy {\bf 180},  165  (1995).

\bibitem{Balcou-Dutriaux:1997}
P. Balcou and L. Dutriaux, Phys. Rev. Let. {\bf 78},  851  (1997).

\bibitem{Jung-Campbell-Chinowsky-Mar-Yee:1998}
L.~D. Jung {\it et~al.}, Langmuir {\bf 14},  5636  (1998).

\bibitem{interfero}
P.R.Berman Ed., Academic Press, Atom Interferometry, London (1997).

\bibitem{CHenkel}
C. Henkel {\it et~al}, Appl. Phys. B {\b 69}, 277 (1999).

\bibitem{Balykin}
V.I.Balykin {\it et~al}, Opt. Commun., {\bf 145}, 322 (1998).

\bibitem{Landragin-Courtois-Labeyrie-Vansteenkiste-Westbrook-Aspect:1996}
A. Landragin {\it et~al.}, Phys. Rev. Lett. {\bf 77},  1464  (1996).

\bibitem{Esslinger-Weidemuller-Hemmerich-Hansch:1993}
T. Esslinger, M. Weidem{\"{u}}ller, A. Hemmerich, and T.~W. H{\"{a}}nsch, Opt.
  Let. {\bf 18},  450  (1993).

\bibitem{Girard-Courjon:1990}
C. Girard and D. Courjon, Phys. Rev. B {\bf 42},  9340  (1990).

\bibitem{Keller-Xiao-Bozhevolnyi:1992}
O. Keller, M. Xiao, and S. Bozhevolnyi, Surf. Sci. {\bf 280},  217  (1992).

\bibitem{Keller:1996}
O. Keller, Physics Reports {\bf 268},  85  (1996).

\bibitem{Girard-Dereux-Weeber:1998}
C. Girard, A. Dereux, and J.~C. Weeber, Phys. Rev. E {\bf 58},  1081  (1998).

\bibitem{Martin-Girard-Dereux:1995:2}
O.~J.~F. Martin, C. Girard, and A. Dereux, J. Opt. Soc. Am. A {\bf 13},  1801
  (1995).

\bibitem{Richard:2001}
N. Richard, Phys. Rev. E {\bf 63},  26602  (2001).

\bibitem{Girard-Bouju-Dereux:1993}
C. Girard, X. Bouju, and A. Dereux,  in {\em Near-Field Optics}, Vol.~E 242 of
  {\em NATO ASI}, edited by D. Pohl and D. Courjon (Klu\-wer, Dor\-drecht,
  1993), pp.\ 199--208.

\bibitem{Greffet-Carminati:1997}
J.-J. Greffet and R. Carminati, Progress in Surface Science {\bf 56},  133
  (1997).

\bibitem{Martin-Piller:1998}
N.~B. Piller and O.~J.~F. Martin, IEEE Trans. Antennas Propag. {\bf 46},  1126
  (1998).

\bibitem{Martin-Girard-Smith-Schultz:1999}
O.~J.~F. Martin, C. Girard, D.~R. Smith, and S. Schultz, Phys. Rev. Let. {\bf
  82},  315  (1999).

\bibitem{Girard-Weeber-Dereux-Martin-Goudonnet:1997}
C. Girard {\it et~al.}, Phys. Rev. B {\bf 55},  16487  (1997).

\bibitem{Girard:1992}
C. Girard, Phys. Rev. B {\bf 45},  1800  (1992).

\bibitem{Palik:1985}
D. Palik, {\em Handbook of Optical Constants of Solids} (Academic Press, New
  York, 1985).

\bibitem{Petit:1980}
R. Petit, {\em Electromagnetic theory of gratings} (Springer Verlag: Topics in
  current physics, Heidelberg, 1980), Vol.~22.

\bibitem{Maystre-Neviere:1978}
D. Maystre and M. Nevi\`ere, J. Opt. {\bf 9},  301  (1978).

\bibitem{Montiel-Neviere:1994}
F. Montiel and M. Neviere, J. Opt. Soc. Am. A {\bf 11},  3241  (1994).

\bibitem{Kobayashi-Sangu-Ito-Ohtsu:2000}
K. Kobayashi, S. Sangu, H. Ito, and M. Ohtsu, Phys. Rev. A {\bf 63},  13806
  (2000).

\bibitem{Birkl-Buchkremer-Dumke-Ertmer:2001}
G. Birkl, F.~B.~J. Buchkremer, R. Dumke, and W. Ertmer, Opt. Com. {\bf 191},
  67  (2001).

\bibitem{Barnes:1998}
W.~L. Barnes, J. Mod. Opt. {\bf 45},  661  (1998).

\bibitem{Dereux-Girard-Weeber:2000}
A. Dereux, C. Girard, and J. Weeber, J. Chem. Phys. {\bf 112},  7775  (2000).

\bibitem{Agarwal:1975c}
G.~S. Agarwal, Phys. Rev. A {\bf 11},  253  (1975).

\bibitem{Wylie-Sipe:1984}
J.~M. Wylie and J.~E. Sipe, Phys. Rev. A {\bf 30},  1185  (1984).

\bibitem{Girard-Martin-Dereux:1995}
C. Girard, O. J. F. Martin, and A. Dereux, Phys. Rev. Lett. {\bf 75}, 3098 (1995)

\bibitem{Metiu:1984}
H. Metiu, Progress in Surface Science {\bf 17},  153  (1984).

\end{thebibliography}
\end{document}